\documentclass[preprintnumbers,amsmath,amssymb]{revtex4}
  \usepackage{graphicx}

\begin{document}
\title{Nonergodic Brownian oscillator: Low-frequency response}
\author{Alex V. Plyukhin}
\email{aplyukhin@anselm.edu}
 \affiliation{ Department of Mathematics,
Saint Anselm College, Manchester, New Hampshire 03102, USA 
}

\date{\today}

\begin{abstract}
An undisturbed  Brownian oscillator  may not reach 
thermal equilibrium with the thermal bath 
 due to the formation of a localized normal mode. The latter may emerge when the spectrum of the thermal bath has a finite upper bound $\omega_0$ and the oscillator natural frequency exceeds a critical value $\omega_c$, which depends on the specific form of the bath spectrum.  
We consider the response of the oscillator with and without a localized mode
to the external periodic force  with frequency $\Omega$ lower than $\omega_0$. The results complement 
those obtained earlier for the high-frequency response at $\Omega\ge \omega_0$ and require a different mathematical approach. The signature property of the high-frequency response is 
resonance when the external force frequency $\Omega$ coincides with the frequency of the localized mode
$\omega_*$.
In the low-frequency 
domain $\Omega<\omega_0$ the condition of resonance 
$\Omega=\omega_*$ cannot be met (since $\omega_*>\omega_0$). Yet, 
in the limits $\omega\to\omega_c$ and $\Omega\to\omega_0^-$,
the oscillator shows a peculiar 
quasi-resonance response with an  amplitude increasing with time sublinearly.
\end{abstract}

\maketitle

\section{Introduction.} 
Consider a dissipative harmonic oscillator, with the mass $m$ and natural frequency $\omega$,
driven by an external ac force $F_{ex}(t)=F_0\sin (\Omega t)$, and
described by the generalized 
Langevin equation~\cite{Zwanzig,Weiss}
\begin{eqnarray}
  m\,\ddot{x}=-m\,\omega^2 x-m\int_0^tK(t-\tau)\,\dot{x}(\tau)\, d\tau
  +F_0\sin(\Omega t)+\xi(t).
\label{GLE}
\end{eqnarray}
The noise $\xi(t)$ is zero-centered, $\langle \xi(t)\rangle=0$, and 
connected with the dissipation kernel $K(t)$ by the standard fluctuation-dissipation relation 
$\langle \xi(t) \xi(t')\rangle=m k_BT K(t-t')$.
Comparing to its Markovian (time-local) counterpart
\begin{eqnarray}
  m\,\ddot{x}=-m\,\omega^2 x-m\,\gamma\,\dot{x}
  +F_0\sin(\Omega t)+\xi(t),
\label{LE}
\end{eqnarray}
the generalized Langevin equation (\ref{GLE}) may hold
on the time scale
comparable  with the relaxation time of the thermal bath.
The latter is important for systems 
with a broad hierarchy of relevant time scales, e.g. for biomolecules
in aqueous solution~\cite{Xie1,Xie2,Goychuk}.

Some types of dissipation kernels $K(t)$ in Eq. (\ref{GLE}) are of particular interest.  
One special type, relevant to a broad class of open systems~\cite{Weiss,Plyukhin_frac},   includes  power-law kernels 
$K(t)\sim t^{-\alpha}$ with  $0<\alpha<1$.  Generalized Langevin equations with such kernels,
and also open systems they describe, are often referred to as fractional.
For fractional Langevin equations 
the integral representing the dissipation coefficient
$\gamma=\int_0^\infty K(t)\,dt$ diverges and therefore the Markovian approximation 
$K(t)\to 2\gamma\delta (t)$ does not exist.
Relaxation and response properties of the fractional oscillator were
studied in~\cite{Barkai1,Barkai2}. Note, however, that there are other definitions
of a fractional oscillator in the literature~\cite{Achar,Stanislav,Ryabov}.

In this paper we consider a kernel $K(t)$ of another special type for which  the spectrum
$\rho(\omega)=\int_0^\infty K(t)\cos(\omega t) dt$ has a finite upper bound $\omega_0$.
The latter can be viewed as a highest frequency of
fictitious  oscillators the thermal bath is comprised of.
Solutions of the generalized Langevin equation with such kernels
may be qualitatively different from those of the corresponding
Markovian equations.
A finite value of the cutoff frequency $\omega_0$ is
known to be a condition for the formation of a localized normal mode
with frequency $\omega_*$ outside of the bath spectrum,  $\omega_*>\omega_0$
\cite{Montroll,Takeno,Kashiwamura,Rubin}.
The localized mode does not exchange energy with the thermal bath and behaves as an
isolated oscillator. As a result,
an open system with a localized mode does not reach thermal
equilibrium  (does not thermalize).
Instead, it evolves into a cyclostationary state where dynamical variables and correlations, averaged
over an ensemble, oscillates in time with frequency of the localized mode $\omega_*$. Cyclostationary states are not stationary and therefore are manifestly nonergodic (ensemble and time averages are not equal). Following Refs.~\cite{Plyukhin,Plyukhin2}, we shall refer to an open oscillator with a localized mode as nonergodic.

Relaxation properties of nonergodic Brownian oscillators 
in the absence of an external driving force ($F_0=0$) 
were studied in~\cite{Kemeny,Dhar,Plyukhin}. Response properties to the external 
harmonic force were addressed in Ref.~\cite{Plyukhin2}
for the case when the external frequency is higher than or equal to the spectrum cutoff frequency, $\Omega\ge\omega_0$.
That domain is of particular interest because it covers  setups
when the external frequency equals the frequency of the localized mode
$\Omega=\omega_*$ (recall that $\omega_*>\omega_0$).  
In that case the system shows unbounded resonance
as for an isolated oscillator with the natural frequency $\omega_*$. 
Another conspicuous regime occurs
when the oscillator natural frequency $\omega$ takes a critical value $\omega_c$
separating ergodic and nonergodic configurations
(without and with a localized mode, respectively).
In that case 
the cutoff frequency $\omega_0$ has the  meaning of the frequency of the
incipient localized mode, $\omega_0=\lim_{\omega\to\omega_c}\omega_*$.
For the setup $(\omega=\omega_c, \,\Omega=\omega_0)$ the oscillator shows
an unusual resonance (``quasi-resonance'') when the amplitude of the response
$q(t)=\langle x(t)\rangle$ increases with time sublinearly.

The goal of this paper is to extend the analysis of Ref.~\cite{Plyukhin2} for $\Omega\ge \omega_0$
to the low-frequency domain $\Omega<\omega_0$. 
Since the localized mode frequency $\omega_*$
is higher than $\omega_0$, the domain $\Omega<\omega_0$ does not include
resonance and quasi-resonance setups mentioned above.   
Still, the low-frequency response may be of interest from both theoretical and experimental points of view. It requires a separate consideration because 
the expressions obtained in Ref.~\cite{Plyukhin2} for $\Omega\ge \omega_0$ involve integrals which diverge for $\Omega<\omega_0$. We will show  that  the results of Ref.~\cite{Plyukhin2} can be  extended for the low-frequency domain merely by defining relevant  improper integrals in the sense of Cauchy principal value. 
That little correction, however, leads to the significant difference 
in the explicit expressions for the low- and high-frequency responses.

In order to make the paper self-consistent, in sections II and III we recapitulate the model and special solutions, including quasi-resonance. 
That sections overlap with Ref.~\cite{Plyukhin2}.
In section IV  we show that the results obtained for the high-frequency response  cannot be directly applied  for the low-frequency domain
and suggest an {\it ad hoc}  remedy for the  extension.
An alternative technique,  addressing the low-frequency response, is  developed in  sections
V and VI. 

\section{Model}

As in Ref.~\cite{Plyukhin2}, we consider a Brownian oscillator described by the generalized
Langevin equation (\ref{GLE}) with a specific dissipation kernel
\begin{eqnarray}
  K(t)=\frac{\omega_0^2}{4}\,[J_0(\omega_0t)+J_2(\omega_0t)]
    =\frac{\omega_0}{2}\,\frac{J_1(\omega_0 t)}{t},
\label{K}
\end{eqnarray}
where $J_n(x)$'s are Bessel functions of the first kind.
The kernel has the absolute maximum at $t=0$ and for $t>0$
it oscillates with an amplitude decaying with time as $t^{-3/2}$.
The kernel's  spectral density $\rho(\nu)$ has the upper bound $\omega_0$,
\begin{eqnarray}
  \rho(\nu)=\int_0^\infty K(t)\cos(\nu t) \,dt=\frac{1}{2}\,\sqrt{\omega_0^2-\nu^2}
  \,\,\theta(\omega_0-\nu).
\label{rho}
\end{eqnarray}
Here $\theta(x)$ is the step function. As was mentioned in the 
Introduction, a finite upper bound of the bath's spectrum 
is a condition of the formation of a localized mode. The frequency of the latter
$\omega_*$ is not explicitly present in Eqs. (\ref{GLE}) and (\ref{K}), and can be viewed
as a hidden parameter of the model.

The generalized Langevin equation with kernel (\ref{K}) describes
dynamics of the terminal  atom of a semi-infinite harmonic chain, 
when the atom
is subjected, in addition to the force from the  neighbour atom,  
to  an external harmonic potential and a time-periodic force. 
The rest of the chain
plays the role of a thermal bath.
The setting can be considered as an extension
of archetypal Rubin's model~\cite{Weiss,Rubin}. Although our choice of the dissipation kernel (\ref{K}) 
is somewhat accidental, it appears to be relevant in a broad context of 
the dynamics of low-dimensional systems with localized modes~\cite{Chalopin}.
One may expect that our model reflects some general features of systems and processes
whose spectral densities have finite upper bounds.


Taking average of Eq. (\ref{GLE}), one gets  
for the average coordinate 
\begin{eqnarray}
q(t)=\langle x(t)\rangle,
\end{eqnarray}
which we shall refer below as the response, 
the deterministic equation
\begin{eqnarray}
\ddot{q}=-\omega^2 q-\int_0^tK(t-\tau)\,\dot{q}(\tau) d\tau
  +\frac{F_0}{m}\sin(\Omega t).
\label{eq_q}
\end{eqnarray}
We assume that the external force starts to act at $t=0$ and that at earlier times the oscillator is in thermal equilibrium. The relevant initial conditions are zero, $q(0)=\dot q(0)=0$. Then the
solution of Eq. (\ref{eq_q}) in the Laplace domain is
\begin{eqnarray}
\tilde q(s)=\frac{1}{m}\,\tilde G(s)\,\tilde F_{ex}(s)
\label{sol_Laplace}
\end{eqnarray}
where the Laplace transforms of the Green's function $G(t)$ and the external force  $F_{ex}(t)=F_0\,\sin(\Omega t)$
are
\begin{eqnarray}
  \tilde G(s)=\frac{1}{s^2+s\,\tilde K(s) +\omega^2}, \quad
     \tilde F_{ex}(s)=F_0\,\frac{\Omega}{s^2+\Omega^2}.
  \label{aux1}
  \end{eqnarray}
 In the time domain the solution has the form of the convolution
\begin{eqnarray}
  q(t)=\frac{F_0}{m}\,\int_0^t G(t-\tau)\,\sin(\Omega\,\tau)\,d\tau.
\label{sol_time}
  \end{eqnarray} 
 One can show, see e.g.~\cite{Plyukhin},  that the Green's function $G(t)$ defined by Eq. (\ref{aux1}) also has the meaning of a 
normalized correlation function  $\langle x(0)\dot x(t)\rangle$.  Then Eq. (\ref{sol_time}) can be viewed as a Kubo relation, which expresses 
a response  in terms of  an equilibrium correlation function.

There are two ways to proceed. In the first way, which we shall follow in Sec. IV, 
the response is evaluated in the time domain using  Eq. (\ref{sol_time}).
This method has the advantage that 
for the given model the Green's function in the time domain $G(t)$ is already known~\cite{Plyukhin}.
In the second way, one finds the response by evaluating  the inverse Laplace transform
of the right-hand side of Eq. (\ref{sol_Laplace}).
Although the two methods are clearly equivalent, 
we shall see that for the low-frequency response at $\Omega<\omega_0$ there are technical reasons 
to prefer one method over the other.

It was shown in ~\cite{Plyukhin} that 
for the given model the Green's function in the time domain $G(t)$ 
takes the form
\begin{eqnarray}
G(t)=
\begin{cases}
    G_e(t), & \text{if $\omega\le\omega_c$},\\
G_e(t)+G_0\,\sin(\omega_* t), & \text{ if $\omega>\omega_c$}.
  \end{cases}
\label{G_gen}
\end{eqnarray}
Here the critical oscillator frequency 
\begin{eqnarray}
  \omega_c=\omega_0/\sqrt{2}\approx 0.707\,\omega_0
\end{eqnarray}
separates ergodic (thermalizing)  configurations with $\omega\le\omega_c$ and nonergodic (non-thermalizing) configurations with $\omega>\omega_c$.
The ergodic component $G_e(t)$ has the same form for $\omega\le \omega_c$ and for $\omega>\omega_c$,
\begin{eqnarray}
G_e(t)=\frac{4}{\pi\omega_0}\int_0^1 \frac{\sin(x\,\omega_0\,t)\,x\,\sqrt{1-x^2}\,dx}
{(1-4\,\lambda^2)\,x^2+4\,\lambda^4},
\label{Ge}
\end{eqnarray}
where $\lambda$ denotes the dimentionless oscillator frequency in units of $\omega_0$,
\begin{eqnarray}
\lambda=\omega/\omega_0.
\end{eqnarray}
Note that in Ref.~\cite{Plyukhin}  $\lambda$ stands for $(\omega/\omega_0)^2$.
For two values of the oscillator frequency, $\omega=\omega_0/2$ and $\omega=\omega_c$, 
Eq. (\ref{Ge}) reduces to more explicit forms in terms Bessel functions, 
see Eqs. (\ref{G3})-(\ref{G4}) below. 

One can show that $G_e(t)$, given by Eq. (\ref{Ge}),  vanishes at long times for all values of $\omega$ (or $\lambda$).
Then for $\omega\le \omega_c$ the Green's function $G(t)=G_e(t)$  goes at long times to zero too. Since $G(t)$ also has the meaning of the normalized correlation
$\langle x(0) \dot x(t)\rangle $~\cite{Plyukhin},  the asymptotic  vanishing of $G(t)$
means that the initial correlations die out and the system  
approaches thermal equilibrium.

Eq. (\ref{G_gen}) shows that the localized mode with frequency $\omega_*$ develops for
$\omega>\omega_c$.
The localized mode amplitude $G_0$ and frequency $\omega_*$
of  the nonergodic component
are given by the following expressions~\cite{Plyukhin}
\begin{eqnarray}
  \omega_0\,G_0=\frac{8\,\lambda^2-4}{(4\,\lambda^2-1)^{3/2}}, \qquad
 \omega_*=\frac{2\,\lambda^2}{\sqrt{4\lambda^2-1}}\,\,\omega_0.
  \label{omega_loc}
\end{eqnarray}
Note that $G_0=0$ and the nonergodic component vanishes for $\omega=\omega_c$ (for $\lambda=\lambda_c=\omega_c/\omega_0=1/\sqrt{2}$).  Therefore 
$q(t)$ as a function of $\omega$ is continuous
including the critical point $\omega=\omega_c$.
For configurations with  $\omega>\omega_c$, with the localized mode developed, the Green's function and initial correlations do not vanish
at long times. Instead of evolving toward thermal equilibrium, the oscillator in the course of time 
reaches the cyclostationary state where the means and correlations oscillate with
the frequency of the localized mode  $\omega_*$.

\section{Special cases and quasi-resonance}

For the two values of the oscillator frequency, $\omega=\omega_0/2$ and
$\omega=\omega_c$,
the inverse transform
of Eq. (\ref{G2}) can be compactly expressed in terms of Bessel functions,
\begin{eqnarray} 
  G(t)&=&\frac{8}{\omega_0^2 t}\,J_2(\omega_0 t)=\frac{2}{\omega_0}\,
  [J_1(\omega_0t)+J_3(\omega_0 t)], \qquad \omega=\omega_0/2,\label{G3}\\
  G(t)&=&\frac{2}{\omega_0}\,J_1(\omega_0 t), \qquad \omega=\omega_c=\omega_0/\sqrt{2}.
  \label{G4}
\end{eqnarray}
These expressions can  also be obtained from Eq. (\ref{Ge})
using 
the integral forms of the Bessel functions.
Both special frequencies $\omega_0/2$ and $\omega_c$ 
correspond to ergodic configurations: The Green's function 
vanishes at long times, 
and the oscillator evolves to  thermal equilibrium.
However, one observes that for $\omega=\omega_c$  the Green's function  decays slower,
which can be viewed  as a precursor of the formation of the localized mode for $\omega>\omega_c$.

Substituting Eqs. (\ref{G3})  and (\ref{G4}) into Eq. (\ref{sol_time}) yields
\begin{eqnarray}
  q(t)&=&
  \frac{2F_0}{m\omega_0}\,\int_0^t
       [J_1(\omega_0\tau)+J_3(\omega_0t)]\,\sin[\Omega\,(t-\tau)]\,dt, 
       \quad\omega=\omega_0/2,
       \label{q1}\\
       q(t)&=&
  \frac{2F_0}{m\omega_0}\,\int_0^t
       J_1(\omega_0\tau)\,\sin[\Omega\,(t-\tau)]\,dt,
       \quad\omega=\omega_c.
       \label{q2}
\end{eqnarray}
We can evaluate these convolution integrals analytically only for $\Omega=\omega_0$.
Of particular interest is the setting $(\omega=\omega_c,\,\Omega=\omega_0)$, when 
Eq. (\ref{q2}) takes the form
\begin{eqnarray}
  q(t)=
  \frac{2F_0}{m\omega_0}\,\int_0^t
       J_1(\omega_0\tau)\,\sin[\omega_0\,(t-\tau)]\,dt,
       \qquad(\omega=\omega_c,\,\Omega=\omega_0).
       \label{q22}
\end{eqnarray}
Taking into account the convolution integral
\begin{eqnarray} 
  \int_0^x J_1(y)\,\sin(x-y)\,dy=\sin(x) -x J_0(x),
  \label{Kapteyn}
\end{eqnarray}
see Ref.~\cite{Watson} and the Appendix in Ref.~\cite{Plyukhin2}, one finds~\cite{Plyukhin2}
\begin{eqnarray} 
  q(t)=\frac{2\,F_0}{m\,\omega_0^2}\,\Big\{\sin(\omega_0 t)
  -\omega_0t\,J_0(\omega_0t)
  \Big\},\qquad
  (\omega=\omega_c,\,\Omega=\omega_0). 
\label{case2}
\end{eqnarray}
Here the first term is the anticipated steady-state solution
oscillating with the frequency of the driving force $\Omega=\omega_0$.
The second term, however, is neither stationary nor transient. 
It oscillates with an amplitude
increasing indefinitely in time as $\sqrt{t}$.
Following Ref. ~\cite{Plyukhin2} we refer to such  
response  as quasi-resonance. 
It can be viewed as 
a resonance between  the external harmonic force 
and the incipient
localized mode in the critical configuration with $\omega=\omega_c$.
Note that according Eq. (\ref{omega_loc}) the cutoff frequency $\omega_0$ also has the meaning of the frequency of the incipient localized mode, $\lim_{\omega\to\omega_c}\omega_*=\lim_{\lambda\to\lambda_c} \omega_*=\omega_0$.

The second configuration, for which the response can be expressed  in a compact form, is 
$(\omega=\omega_0/2,\,\Omega=\omega_0)$. In that case 
Eq. (\ref{q1}) reads
\begin{eqnarray}
  q(t)=
  \frac{2F_0}{m\omega_0}\,\int_0^t
       [J_1(\omega_0\tau)+J_3(\omega_0t)]\,\sin[\omega_0\,(t-\tau)]\,dt.
       \label{q11}
\end{eqnarray}
Taking into account integral (\ref{Kapteyn}) and its generalization~\cite{Bailey}  
\begin{eqnarray} 
\int_0^t J_3(y)\,\sin(x-y)\,dy=-x \, J_2(x)+6J_1(x)-3\sin(x)=x \, J_0(x)+4J_1(x)-3\sin(x)
\end{eqnarray}
(here in the second equality we use the recurrence relation $\frac{2n}{x} J_n=J_{n-1} + J_{n+1}$  with $n=1$), yields 
\begin{eqnarray} 
  q(t)=\frac{4\,F_0}{m\,\omega_0^2}\,\Big\{
\sin(\omega_0 t-\pi)+2J_1(\omega_0t)\Big\}, \qquad 
  (\omega=\omega_0/2,\,\Omega=\omega_0). 
\label{case1}
\end{eqnarray}
The structure of this solution is similar to that
for the normal damped and fractional oscillators.
The first term is the steady-state solution which oscillates with the frequency of the external
force $\Omega=\omega_0$. 
The second  term is a transient which vanishes at long times.
One notable feature is a slow decay of the transient term;
its amplitude decays as $1/\sqrt{t}$. Another remarkable feature is that 
the phase shift $\pi$ of the steady-state response is the same as for an isolated (undamped) 
oscillator when $\Omega>\omega$~\cite{French}.
Recall that  
for the normal damped oscillator, described by Eq. (\ref{LE}), the phase shift reaches the value $\pi$ only in the limit $\Omega\to\infty$~\cite{French}.

For $\Omega\ne\omega_0$ the integral
expressions (\ref{q1}) and (\ref{q2}) 
apparently cannot be further simplified and may be used to evaluate the response numerically. In that way, 
for the ergodic configuration with $\omega=\omega_0/2$ we find that $q(t)$
evolves into a steady-state form
\begin{eqnarray} 
  q(t)\to A\,\sin (\Omega t-\delta)
\label{steady}
\end{eqnarray} 
with the  amplitude $A>0$ weakly decreasing  with increasing  $\Omega$, 
and the phase shift $\delta$ monotonically increasing from $0$ for small $\Omega$ 
to $\pi$ for $\Omega\gg\omega$.
There is no resonance when $\Omega$ is close to $\omega$, and 
on the whole, the response is qualitatively similar 
to that of the  overdamped normal damped oscillator, described by 
Eq. (\ref{LE}) with $\gamma>\omega$~\cite{French}.

For the critical configuration with $\omega=\omega_c$, the response, described by Eq. (\ref{q2}), may show a variety of patterns. 
For $\Omega=\omega_0$ the response has the quasi-resonance form (\ref{case2}) and never reaches
the steady state (\ref{steady}). When  $|\Omega/\omega_0-1|$ is finite but small, the response  evolves at long times
to the steady state (\ref{steady}), but at shorter times shows beats.  The duration and amplitude of the first beat increase  when $\Omega$ is getting closer to $\omega_0$ and diverge in the limit $\Omega\to\omega_0$.
When $\Omega$ is not close to $\omega_0$  the response quickly reaches the steady state form (\ref{steady}),
where  the amplitude  weakly depends on $\Omega$ for $\Omega<\omega_0$ and decreases with increasing $\Omega$ when
$\Omega>\omega_0$.

\section{Ad hoc approach}
Consider general ergodic configurations with $\omega\le \omega_c$. 
The localized mode is absent,
the Green's function has only  ergodic component, $G(t)=G_e(t)$,  and
the response, according to Eq. (\ref{sol_time}), is 
\begin{eqnarray}
  q(t)=
  \frac{F_0}{m}\,\int_0^t G_e(t-\tau)\,\sin(\Omega\,\tau)\,d\tau.
\label{sol_time2}
\end{eqnarray}
Substituting Eq. (\ref{Ge}) for $G_e(t)$ into Eq. (\ref{sol_time2}),
changing the integration order, 
and integrating over $\tau$ yields
\begin{eqnarray}
  q(t)=\frac{F_0}{m\omega_0^2}\Big\{
  B\,\sin(\Omega t)+\varphi(t)\Big\},
\label{qe}
\end{eqnarray}
where the dimensionless amplitude $B$ and the transient $\varphi(t)$ are
\begin{eqnarray}
 B&=&-\frac{4}{\pi}
  \int_0^1\frac{x^2\sqrt{1-x^2}\,dx}
      {
    \Big[(1-4\lambda^2)\,x^2+4\lambda^4\Big]
   \left(\Lambda^2-x^2\right) 
      },
      \label{amp}\\
 \varphi(t)&=&\frac{4\Lambda}{\pi}
  \int_0^1\frac{x\sqrt{1-x^2}\,\sin(x\,\omega_0\,t)\,dx}
      {
    \Big[(1-4\lambda^2)\,x^2+4\lambda^4\Big]
   \left(\Lambda^2-x^2\right) 
      }.
      \label{varphi}
\end{eqnarray}
In these expressions  $\lambda$ and $\Lambda$ denote the dimensionless oscillator and external force frequencies
in units of the cutoff frequency $\omega_0$,
\begin{eqnarray}
\lambda=\omega/\omega_0, \qquad \Lambda=\Omega/\omega_0. 
\end{eqnarray}
One observes that 
Eqs. (\ref{qe})-(\ref{varphi}) make sense, and have been used in Ref.~\cite{Plyukhin2}, 
only for the high-frequency response at $\Omega\ge \omega_0$, or $\Lambda\ge 1$.
In the low-frequency domain $\Omega<\omega_0$, or $\Lambda<1$, 
the integrands in Eqs. 
(\ref{amp}) and (\ref{varphi}) are singular at $x=\Lambda$ and the integrals diverge.

Recall that in deriving the above  equations we substituted integral (\ref{Ge}) over $x$  into 
integral (\ref{sol_time2}) over $\tau$ and  then, in a rather cavalier  manner,
changed the integration order $\int\!d\tau\int\!dx\to \int\!dx\int\!d\tau$. 
The conditions of that last operation, regulated by Fubini's theorem, are not satisfied for $\Lambda<1$ 
since the integrand is not continuous  in the integration domain.

However, one notices that the integrals (\ref{amp}) and (\ref{varphi}) converge in Cauchy's principal value sense:
\begin{eqnarray}
  \text{p.v.}\int_0^1f(x)\,dx=\lim_{\epsilon\to 0^+}\left\{
  \int_0^{\Lambda-\epsilon} f(x)\,dx+\int_{\Lambda+\epsilon}^1 f(x)\,dx
  \right\}.
  \label{cpv}
\end{eqnarray}
This suggests that 
Eqs. (\ref{qe})-(\ref{varphi}) are valid for the low-frequency response as well,  
if one interpret the integral expressions for $B$ and $\varphi(t)$ 
in the principal value sense,
\begin{eqnarray}
 B&=&-\frac{4}{\pi}\,
 \text{p.v.}
  \int_0^1\frac{x^2\sqrt{1-x^2}\,dx}
      {
    \Big[(1-4\lambda^2)\,x^2+4\lambda^4\Big]
   \left(\Lambda^2-x^2\right) 
      },
      \label{B4}\\
 \varphi(t)&=&\frac{4\Lambda}{\pi}\,
 \text{p.v.}
  \int_0^1\frac{x\sqrt{1-x^2}\,\sin(x\,\omega_0\,t)\,dx}
      {
    \Big[(1-4\lambda^2)\,x^2+4\lambda^4\Big]
   \left(\Lambda^2-x^2\right) 
      }.
      \label{varphi4}
\end{eqnarray}
Whereas this guess seems reasonable, it is rather {\it ad hoc}. In the next section we will  justify Eqs. (\ref{B4}) and (\ref{varphi4})
with another method, briefly mentioned at the end of Sec. II. In that method the 
principal value of the integral for $\varphi(t)$ emerges naturally, 
and the expression for $B$ has a more explicit form, see Eqs. (\ref{result_B}) and (\ref{result_varphi}) below.

\section{Response of ergodic configurations}
An alternative method to evaluate 
the response is to start with Eq. (\ref{sol_Laplace}) for its Laplace transform
$\tilde q(s)$ and to perform the inversion $q(t)= \frac{1}{2\pi i}\int_\text{Br} e^{st} \tilde q(s)\,ds$,
which gives
\begin{eqnarray}
  q(t)= 
  \frac{F_0\Omega}{m}
  \frac{1}{2\pi i}\,
  \int_\text{Br} \frac{e^{st}\,\tilde G(s)\,ds}{(s+i\Omega)(s-i\Omega)}.
  \label{BI}
\end{eqnarray}
Here $\text{Br}$ denotes the Bromwich path in the complex plane 
(a vertical line to the right of all singularities of the 
integrand) and  the transform of the Green's function is
\begin{eqnarray}
\tilde G(s)=\frac{2}{s^2+s\,\sqrt{s^2+\omega_0^2}+2\,\omega^2}.
\label{G2}
\end{eqnarray}
This expression  follows from Eq. (\ref{aux1}) and 
the Laplace transform of the kernel (\ref{K}) 
\begin{eqnarray}
  \tilde K(s)=\frac{1}{2}\, \left(
  \sqrt{s^2+\omega_0^2}-s\right).
\end{eqnarray}
 As shown in Ref.~\cite{Plyukhin},
for ergodic configurations with $\omega\le\omega_c$  
the only singularities of the Green's function $\tilde G(s)$ 
are two branch points at $\pm i\omega_0$.
Two other singularities of the integrand in Eq. (\ref{BI})
are simple poles at $\pm i\Omega$.
Since all singular points are on the imaginary axis,  contour $\text{Br}$  in Eq. (\ref{BI}) 
is an arbitrary vertical line to the right of the origin.

The  evaluation of  the integral (\ref{BI}) can be performed
with the standard technique based on the Cauchy's integral and residue theorems,
but there are two subtleties. The first is that the poles lie on the branch cut, the second is that $\tilde G(s)$ has two branches. Below we outline the main steps of the evaluation.

The first step is to close the Bromwich contour as shown in Fig. 1 (left) and to consider the integral $I$ over the resulting  closed contour $\Gamma$
\begin{eqnarray}
  I= 
  \frac{F_0\Omega}{m}
  \frac{1}{2\pi i}\,
  \int_\Gamma \frac{e^{st}\,\tilde G(s)\,ds}{(s+i\Omega)(s-i\Omega)},
  \label{I}
\end{eqnarray}
assuming 
the limit when the radius $R$
of the large arc
is infinitely large, $R\to\infty$,
and the radii $\epsilon$ of the small (semi-) circles about the branch points $\pm i\omega_0$
and poles $\pm i\Omega$ are infinitely small, $\epsilon\to 0$.
The contribution to $I$ from the arc vanishes, the contributions from the
line above and below the real axis cancel each other, and the contribution 
from the rightmost vertical line equals $q(t)$. Then we get
\begin{eqnarray}
  I=q(t)+I_C,
  \label{I2}
\end{eqnarray}
where $I_C$
is the contribution from the contour 
$C$ depicted in Fig. 1 (right). On the other hand, since there  are no singular points
in the interior of $\Gamma$, Cauchy's integral theorem asserts that $I=0$, and therefore
\begin{eqnarray}
  q(t)=-I_C.
\label{aux2} 
\end{eqnarray}

\begin{figure}[t]
\includegraphics[height=8cm]{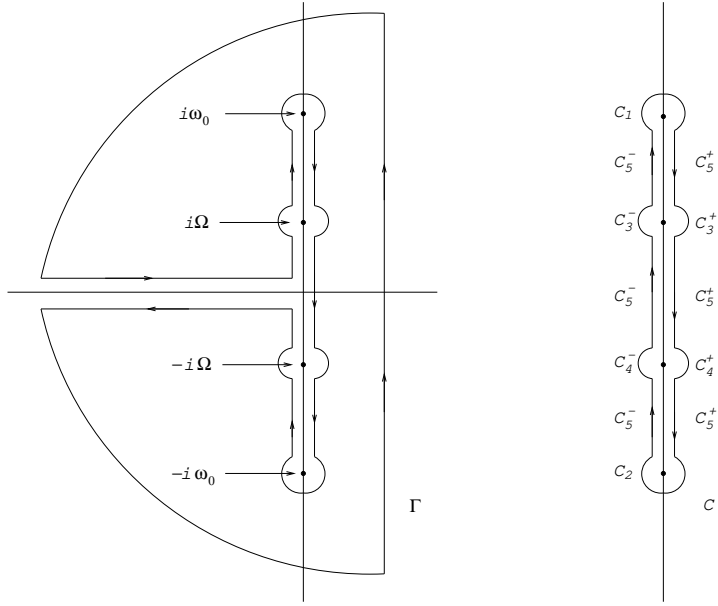}
\caption{Left: Integration contour $\Gamma$ for integral $I$ in Eq. (\ref{I}). Right: Integration contour $C$ for integral $I_C$ in Eqs. (\ref{I2}) and (\ref{IC})
}
\end{figure}

To proceed, 
one has to take into account 
that the Green's function (\ref{G2}) involves 
the two-branch function $f(s)=\sqrt{s^2+\omega^2}$. To choose the branch which is 
physically meaningful, one may recall
that the Green's function $G(t)$ also determines
the equilibrium correlation $\langle q(0)\dot q(t)\rangle$~\cite{Plyukhin} 
\begin{eqnarray}
\langle q(0)\dot q(t)\rangle=\frac{k_BT}{m}\,G(t).
\end{eqnarray}
That implies the initial condition $G(0)=0$, which in the Laplace domain has the form
 \begin{eqnarray}
\lim_{s\to\infty} s\,\tilde G(s)=0.
\label{property}
\end{eqnarray}
One can show that this condition is only satisfied when $\tilde G(s)$ involves the branch of  
$f(s)=\sqrt{s^2+\omega^2}$  which takes real positive (negative) values when $s$ 
is real and positive (negative). 
Explicitly, the physical branch of the function 
\begin{eqnarray}
f(s)=\sqrt{s^2+\omega_0^2}=\sqrt{
(s-i\omega_0)(s+i\omega_0)}
\end{eqnarray}
is convenient to describe using a pair of polar coordinates $(r_1,\theta_1)$ and $(r_2,\theta_2)$  
with the origins
at $s=i\omega_0$ and $s=-i\omega_0$, respectively:
\begin{eqnarray}
f(s)=\sqrt{r_1r_2}\,\exp\left(i\,\frac{\theta_1+\theta_2}{2}\right).
\label{f1}
\end{eqnarray}
For the physical branch the polar angles may be chosen from the same interval
\begin{eqnarray}
-\frac{3\pi}{2}<\theta_1\le \frac{\pi}{2}, \quad -\frac{3\pi}{2}<\theta_2\le \frac{\pi}{2}.
\label{f2}
\end{eqnarray}
For instance, for $s=x$ being real and positive we get $\theta_1+\theta_2=0$, then $f(s)=\sqrt{r_1r_2}$ is real and positive.

To evaluate the integral
\begin{eqnarray}
  I_C=
  \frac{F_0\Omega}{m}
  \frac{1}{2\pi i}\,
  \int_C \frac{e^{st}\,\tilde G(s)\,ds}{(s+i\Omega)(s-i\Omega)}
  \label{IC}
\end{eqnarray}
we 
split it  into five contributions
\begin{eqnarray}
I_C=I_1+I_2+I_3+I_4+I_5,
\end{eqnarray}
corresponding to the regions $C_i$ of the integration path, see Fig. 1 (right).
Contributions $I_1$ and $I_2$ come from the integration
over circles $C_1$ and $C_2$ of radius $\epsilon$ 
about the branch points $\pm i\omega_0$. 
One finds that the integrals are of order of $\epsilon$ for $\omega\ne\omega_c$, and  
of order of $\sqrt{\epsilon}$ for $\omega=\omega_c$.  Thus $I_1$ and $I_2$ for any $\omega$ 
vanish when $\epsilon\to 0$, 
\begin{eqnarray}
I_1=I_2=0.
\end{eqnarray}

Next consider the integrals $I_3, I_4$ over the semi-circles $C_3^\pm,C_4^\pm$ of radius $\epsilon$ about $s=\pm i\Omega$.
Using Eqs. (\ref{f1}) and (\ref{f2}), for $s\in C_3^\pm,C_4^\pm$ in the limit $\epsilon\to 0$ 
one finds
\begin{eqnarray}
f(s)=
\pm \sqrt{\omega_0^2-\Omega^2},
\label{f3}
\end{eqnarray}
where the sign is positive for the right semi-circles  and negative for left semi-circles.
The Green's function (\ref{G2}) for $s\in C_3^\pm, C_4^\pm$ takes the form
\begin{eqnarray}
  \tilde G(s)=\frac{2}
  {2\omega^2-\Omega^2\pm i\Omega\,\sqrt{\omega_0^2-\Omega^2}},
\end{eqnarray}
where again the plus and minus in the denominator correspond
to the right and left semi-circles, respectively. 
Then the straightforward evaluation gives
\begin{eqnarray}
  I_3+I_4=-\frac{F_0}{m}\,
  \frac{2(2\omega^2-\Omega^2)}{4\omega^4+\Omega^2(\omega_0^2-4\omega^2)}\,\sin(\Omega t).
\end{eqnarray}

The last contribution $I_5=I_5^++I_5^-$ comes from the vertical segments $C_5^+$ and $C_5^-$ 
along  the right and left shores of the
branch cut, excluding the $2\epsilon$-wide intervals about $\pm i\Omega$.
In that case 
$s=iy\pm\delta$, where $\delta$ is infinitesimal and 
\begin{eqnarray}
y\in(-\omega_0+\epsilon,-\Omega-\epsilon)\cup(-\Omega+\epsilon, \Omega-\epsilon)\cup(\Omega+\epsilon,\omega_0-\epsilon).
\label{y_region}
\end{eqnarray}
For  $s\in C_5^\pm$ one finds $f(s)=\sqrt{s^2+\omega_0^2}=\pm\sqrt{\omega_0^2-y^2}$, and the Green's function (\ref{G2})
takes the form
\begin{eqnarray}
 \tilde G(s)=\frac{2}{2\omega^2-y^2\pm i y\sqrt{\omega_0^2-y^2}},
\end{eqnarray}
where the plus and minus correspond to $C_5^+$ and $C_5^-$,
respectively. Taking that into account we obtain
\begin{eqnarray}
  I_5=-\frac{F_0\,\Omega}{m}\,\frac{4}{\pi}\, \text{p.v.} \int_0^{\omega_0}
  \frac{y\,\sqrt{\omega_0^2-y^2}\,\sin(yt)\,dy}
  {[4\omega^4+y^2(\omega_0^2-4\omega^2)] \,(\Omega^2-y^2)}.
  \label{I5}
\end{eqnarray}
The principal value of the integral emerges here
quite naturally
since the integration region (\ref{y_region})
has symmetric gaps about $\pm \Omega$.

Summing up contributions yields
\begin{eqnarray}
I_C&=&-\frac{F_0}{m\omega_0^2}\left\{
B\,\sin(\Omega t)+\varphi(t)
\right\}, \label{IC2}\\
B&=&\frac{2\omega_0^2(2\omega^2-\Omega^2)}{4\omega^4+
(\omega_0^2-4\omega^2)\Omega^2},\\
\varphi(t)&=&\frac{4\Omega\,\omega_0^2}{\pi}\,\text{p.v.} \int_0^{\omega_0}
  \frac{y\,\sqrt{\omega_0^2-y^2}\,\sin(yt)\,dy}
  {[4\omega^4+y^2(\omega_0^2-4\omega^2)] \,(\Omega^2-y^2)} .
\end{eqnarray}
Finally, using Eq. (\ref{aux2}) and expressing $B$ and $\varphi(t)$ in terms of dimensionless frequencies $\lambda=\omega/\omega_0$ and $\Lambda=\Omega/\omega_0$ we get
\begin{eqnarray}
q(t)&=&\frac{F_0}{m\omega_0^2}\Big\{
B\,\sin(\Omega t)+\varphi(t)\Big\},
\label{result_q}\\
B&=&\frac{2\,(2\lambda^2-\Lambda^2)}
  {4\lambda^4+(1-4\lambda^2)\,\Lambda^2},\label{B5}
  \label{result_B}\\
\varphi(t)&=&
\frac{4\Lambda}{\pi}\, \text{p.v.} \int_0^{1}
  \frac{x\,\sqrt{1-x^2}\,\sin(x\omega_0 t) \,dx}
  {[4\lambda^4+x^2(1-4\lambda^2)]\,(\Lambda^2-x^2)}.
\label{result_varphi}
\end{eqnarray}
These results, which hold for $\omega<\omega_c$ and $\Omega<\omega_0$ (or,  for $ \lambda<\lambda_c$ and $\Lambda<1$),  are equivalent to  Eqs. (\ref{B4}) and (\ref{varphi4})
obtained in the previous section with method I, but Eq. (\ref{B5}) for $B$ is more explicit,
and  the principle value in the integral expression for $\varphi(t)$
is  now totally justified, see the remark after Eq. (\ref{I5}).

\begin{figure}[t]
  \includegraphics[
    width=10cm,
    height=9cm]{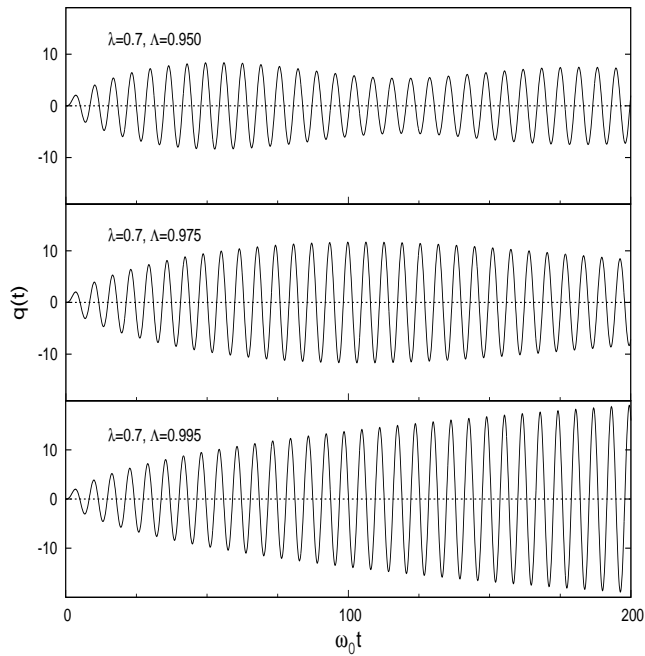}
\caption{The average coordinate $q(t)$, in units of $q_0=F_0/(m\omega_0^2)$,
  for ergodic configurations with the dimensionless oscillator
  frequency $\lambda=\omega/\omega_0=0.7$ (just below the critical
  value $\lambda_c=1/\sqrt{2}\approx 0.707$) 
and three values of the dimensionless external frequency 
$\Lambda=\Omega/\omega_0$, evaluated with Eqs. (\ref{result_q})-(\ref{result_varphi}).
At much longer times the beat pattern dies out and
$q(t)$ takes the steady-state form Eq. (\ref{steady2}).  
The plot at the bottom is close to that
for the quasi-resonance solution Eq. (\ref{case2}). 
The integral in Eq. (\ref{result_varphi}) is evaluated numerically with the {\it NIntegration} routine of {\it Wolfram Mathematica} using a principal value option~\cite{Mathematica}.} 
\end{figure}

One can verify numerically that the function $\varphi(t)$ defined by Eq. (\ref{result_varphi}) evolves at long times 
into a harmonic form oscillating with frequency $\Omega$. Therefore, at long times both terms in Eq. (\ref{result_q})
oscillate with frequency $\Omega$, and so does the total solution,
\begin{eqnarray}
q(t)\to A\,\sin(\Omega t-\delta).
\label{steady2}
\end{eqnarray}
Here the amplitude $A$ and the phase shift $\delta$ depend on $\omega$ and $\Omega$ (or, on $\lambda$ and $\Lambda$) and are
difficult to express in an analytical forms.
The response relaxes to the steady state form (\ref{steady2}) on the time scale which increases and, in fact, diverges, when $\omega$ and $\Omega$ are getting close to $\omega_c$ and $\omega_0$, respectively.
For such settings, the solution on a long initial time interval shows beat patterns, see Fig. 2. 
For the critical configuration $\omega=\omega_c$ the duration and amplitude of the first beat
indefinitely increases as $\Omega\to\omega_0^-$, and the solution converges to the quasi-resonance 
form (\ref{case2}). Such behavior is similar (though not numerically identical) to that for the high-frequency response of the critical configuration
($\omega=\omega_c$) when  $\Omega$ approaches $\omega_0$ from the above, $\Omega\to\omega_0^+$, see Fig. 1 in  Ref.~\cite{Plyukhin2}.

\section{Response of nonergodic configurations} 
Consider nonergodic configurations with $\omega>\omega_c$.
The Green's function has now both ergodic and nonergodic (periodic) components,
$G(t)=G_e(t)+G_0\,\sin(\omega_* t)$. We can follow the same approach as in the previous section, but taking into account that 
the Green's function $\tilde G(s)$ in the Laplace domain now 
has two extra poles at 
$\pm i\omega_*$. Since $\omega_*>\omega_0$, 
the integral $I$ given by Eq. (\ref{I}) is not zero but equals to the sum of residues at $s=\pm i\omega_*$,
\begin{eqnarray}
I=Res[e^{st}\,\tilde q(s), i\omega_*]+Res[e^{st}\,\tilde q(s), -i\omega_*],
\end{eqnarray}
where 
\begin{eqnarray}
\tilde q(s)=\frac{1}{m}\,\tilde G(s)\,\tilde F_{ex}(s)=
\frac{1}{m}\,\tilde G(s)\,\frac{\Omega}{s^2+\Omega^2}.
\label{sol_Laplace2}
\end{eqnarray}
On the other hand $I=q(t)+I_C$, see Eq. (\ref{I2}),
and therefore
\begin{eqnarray}
q(t)=-I_C+Res[e^{st}\,\tilde q(s), i\omega_*]+Res[e^{st}\,\tilde q(s), -i\omega_*].
\end{eqnarray}
Here the integral $I_C$ is still given by Eq. (\ref{IC2})
and the  sum of residues can be written as 
\begin{eqnarray}
  Res[e^{st} \tilde q(s),
    i\omega_*]+Res[e^{st}\tilde q(s),-i\omega_*]
=\frac{F_0\Omega}{m\,(\Omega^2-\omega_*^2)}\,
\Big\{Res[e^{st} \tilde G(s), i\omega_*]+
 Res[e^{st} \tilde G(s), i\omega_*]
\Big\}.
\end{eqnarray}
The sum of residues of the Green's function can be readily
evaluated (see Appendix B in~\cite{Plyukhin} for $\mu=1$), or one may just notice that 
it gives the 
nonergodic part of the Green's function $G_0\sin(\omega_* t)$,
which gives
\begin{eqnarray}
  Res[e^{st} \tilde q(s),
    i\omega_*]+Res[e^{st}\tilde q(s),-i\omega_*]
=\frac{F_0\Omega}{m\,(\Omega^2-\omega_*^2)}\,G_0\,\sin(\omega_*t).
\label{residue_sum}
\end{eqnarray}
Combining the results and expressing in terms of dimensionless frequencies, 
we obtain  
\begin{eqnarray}
q(t)&=&\frac{F_0}{m\omega_0^2}\Big\{
B\sin(\Omega t)+C\,\sin(\omega_*t)+\varphi(t)\Big\}, 
\label{result2_q}\\
B&=&\frac{2\,(2\lambda^2-\Lambda^2)}
  {4\lambda^4+(1-4\lambda^2)\,\Lambda^2},
  \label{result2_B}\\
C&=&\omega_0G_0\,\frac{\Lambda}{\Lambda^2-\lambda_*^2}=
-\frac{4\Lambda(2\,\lambda^2-1)}
{\sqrt{4\,\lambda^2-1}\,(4\lambda^4+\Lambda^2-4\lambda^2\Lambda^2
)}
,
\label{result2_C}\\
\varphi(t)&=&
\frac{4\Lambda}{\pi}\, \text{p.v.} \int_0^{1}
  \frac{x\,\sqrt{1-x^2}\,\sin(x\omega_0 t) \,dx}
  {(\Lambda^2-x^2)[4\lambda^4+x^2(1-4\lambda^2)]}.
\label{result2_varphi}
\end{eqnarray}
Here, in the first equation for $C$, $\lambda_*$ stands for the dimensionless localized frequency
$\lambda_*=\omega_*/\omega_0$ and $G_0$ is the amplitude of the nonergodic part of the Green's function, both are given by Eq. (\ref{omega_loc}).

\begin{figure}[t]
  \includegraphics[
    width=10cm,
    height=9cm]{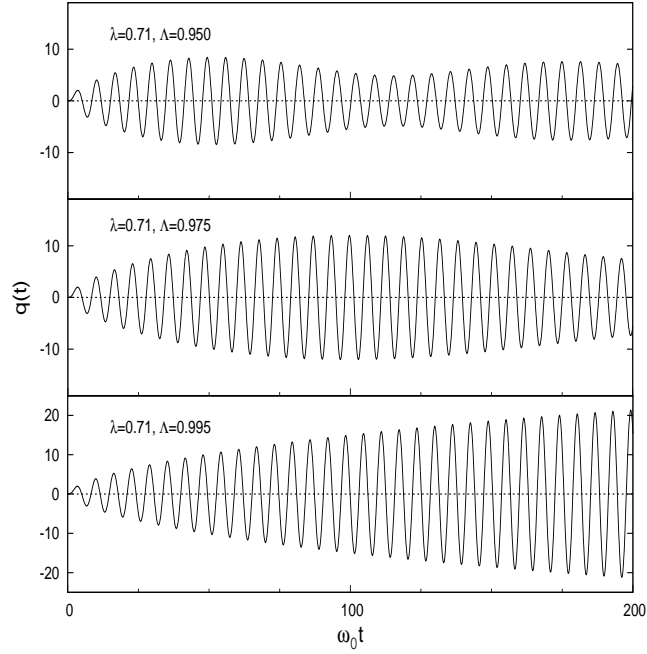}
\caption{The average coordinate $q(t)$, in units of $q_0=F_0/(m\omega_0^2)$,
  for nonergodic configurations with the dimensionless oscillator
  frequency $\lambda=\omega/\omega_0=0.71$ (just above the critical
  value $\lambda_c=1/\sqrt{2}\approx 0.707$) 
and three values of the dimensionless external frequency 
$\Lambda=\Omega/\omega_0$, evaluated with Eqs. (\ref{result2_q})-(\ref{result2_varphi}).
At longer times $q(t)$ shows beats which do not die out 
and continue indefinitely. 
The plot at the bottom is close to that for the quasi-resonance solution Eq. (\ref{case2}). } 
\end{figure}

Compared to the response of ergodic configurations, solution (\ref{result2_q}) involves the additional term
$\frac{F_0}{m\omega_0^2}\,C\,\sin(\omega_*t)$. The term 
$\varphi(t)$ at long times still oscillates with frequency $\Omega$. 
Therefore, the asymptotic long-time response of nonergodic configurations does not oscillate with the frequency of the external force $\Omega$   but has the form of the superposition of two harmonic terms with 
frequencies $\Omega$ and $\omega_*$,
\begin{eqnarray}
q(t)\to A\,\sin(\Omega t-\delta)+C\,\sin(\omega_* t),
\label{steady3}
\end{eqnarray}
In the  considered domain $(\omega>\omega_c, \, \Omega<\omega_0)$ the frequencies
$\omega_*$ and $\Omega$ are separated, $\omega_*>\omega_0$ and $\Omega<\omega_0$,
yet may be close. In that cases the asymptotic solution (\ref{steady3}) 
shows beats which, in contrast to ergodic configurations, 
do not vanish at long times but continue indefinitely.
When $\omega\to\omega_c^+$ and $\Omega\to \omega_0^-$ the response 
converges to the quasi-resonance form (\ref{case2}), see Fig. 3.

\section{Conclusion}
The nonergodic Brownian oscillator considered in this paper is a quite unusual open system. Its  signature properties are the luck of thermalization (when unperturbed)
and unbounded resonance and quasi-resonance response
when the frequency of the external force $\Omega$ coincides with the frequency of the localized mode $\omega_*$. In this paper we considered the low-frequency 
response at $\Omega<\omega_0$, where $\omega_0$ is the upper bound of the thermal bath's spectrum.  In that case the two frequencies $\Omega$ and $\omega_*$
are separated, $\Omega<\omega_0$ and $\omega_*\ge \omega_0$,  and the condition of resonance cannot be met.
However, for the critical and near-to-critical configurations ($\omega$ is equal or close to $\omega_c$) the localized frequency $\omega_*$ is
close to $\omega_0$, and  the frequencies $\Omega$ and $\omega_*$ may be close even for the low-frequency domain $\Omega<\omega_0$.  In such case
the low-frequency response may show a quasi-resonance response, with the amplitude increasing with time sublinearly,   on a finite  but long time interval.

Comparing the results with those obtained in Ref.~\cite{Plyukhin}, 
we found the response 
to be a continuous function of the external frequency, including the point $\Omega=\omega_0$.
Although for $\Omega<\omega_0$ and for $\Omega\ge\omega_0$ the response is given by different expressions, they match at $\Omega=\omega_0$.



\begin{thebibliography}{99}

\bibitem{Zwanzig} R. Zwanzig, {\it Nonequilibrium Statistical Mechanics},
Oxford University Press, NY (2001).

  
\bibitem{Weiss} U. Weiss, {\it Quantum Dissipative Systems}, World Scientific, Singapore (2008).


\bibitem{Xie1} S. C. Kou and X. S. Xie, { Generalized Langevin equation with fractional Gaussian noise: Subdiffusion within a single protein molecule}, Phys. Rev. Lett. 93, 180603 (2004).

\bibitem{Xie2} W. Min, G. Luo, B. J. Cherayil, S. C. Kou, and X. S. Xie,
{\it Observation of a power-law memory kernel for fluctuations within a single protein molecule}, 
Phys. Rev. Lett. 94, 198302 (2005).


\bibitem{Goychuk} I. Goychuk, {\it Viscoelastic subdiffusion: Generalized Langevin equation approach},
Adv. Chem. Phys. 150, 187 (2012).


\bibitem{Plyukhin_frac} A.V. Plyukhin, 
{\it Fractional Langevin equation from damped bath dynamics}, Phys. Rev. E 
99, 052125 (2019).

\bibitem{Barkai1} S. Burov and E. Barkai,
    {\it Critical exponent of the fractional Langevin equation},
    Phys. Rev. Lett. 100, 070601 (2008).
    
    
  \bibitem{Barkai2}  S. Burov and E. Barkai, {\it
    Fractional Langevin equation: Overdamped, underdamped and critical behaviors},
    Phys. Rev. E 78, 031112 (2008).
    


  \bibitem{Achar} B. N. N. Achar, J. W. Hanneken, T. Enck, and T. Clarke,
    {\it Dynamics of the fractional oscillator}, Physica A 297, 361 (2001).
    
  \bibitem{Ryabov} Ya. E. Ryabov and A. Puzenko,
    {\it Damped oscillations in view of the fractional oscillator equation},
    Phys. Rev. E 66, 184201 (2002).

  \bibitem{Stanislav}  A. A. Stanislavsky, {\it Fractional oscillator},
    Phys. Rev. E 70, 051103 (2004).

  
    
\bibitem{Montroll} E. W. Montroll   and R. B. Potts,
{\it Effect of defects on lattice vibrations}, 
Phys. Rev. 100, 525 (1955). 


\bibitem{Takeno} E. Teramoto and S. Takeno, 
{\it Time dependent problems of the localized lattice vibration}, 
Prog. Theor. Phys. 24, 1349, 1960. 



\bibitem{Kashiwamura} S. Kashiwamura, 
{\it Statistical dynamical behaviors of a one-dimensional lattice with an isotopic impurity},
Prog. Theor. Phys. 27, 571, 1962. 


\bibitem{Rubin} R. Rubin, 
{\it Momentum autocorrelation functions and energy transport in harmonic crystals containing isotopic defects},  
Phys. Rev. 131, 964, 1963. 





\bibitem{Kemeny}  G. Kemeny, S. D. Mahanti, and T. A. Kaplan,
  {\it Generalized Langevin equation for an oscillator}, Phys. Rev. B 34, 6288 (1986).


\bibitem{Dhar} A. Dhar and K. Wagh, {\it Equilibration problem for the generalized Langevin equation},
  Europhys. Lett 79, 60003 (2007).

  
  
  \bibitem{Plyukhin} A. V. Plyukhin, {\it Nonergodic Brownian oscillator}, Phys. Rev. E 105, 014121 (2022).



\bibitem{Plyukhin2} A. V. Plyukhin, {\it Nonergodic Brownian oscillator: High-frequency response}, Phys. Rev. E 107, 044107 (2023). 

\bibitem{Chalopin} Y. Chalopin and J. Sparfel,
 {\it Energy bilocalization effect and the emergence of molecular functions in proteins,
 }  Front. Mol. Biosci. 8, 736376 (2021).

\bibitem{French} A. P. French, {\it Vibrations and Waves}, W. W. Norton \& Company, New York, 1971.

    

    
  \bibitem{Watson}
G. N. Watson,
{\it A Treatise on the Theory of Bessel Functions}, second ed.,
Cambridge University Press, London,
1944, section 12.21.


\bibitem{Bailey} W. N. Bailey, {\it Some integrals of Kapteyn's type involving Bessel functions}, 
  Proc. London Math. Soc., s2-30, 422, (1930).



\bibitem{Mathematica} Wolfram Research, Inc., Mathematica, Version 13.2, Champaign, IL (2022).





\end{thebibliography}
\end{document}